\documentclass[conference, 10pt]{IEEEtran}
\IEEEoverridecommandlockouts
\usepackage[compress]{cite}
\usepackage{amsmath,amssymb,amsfonts}
\usepackage{algorithmic}
\usepackage{graphicx}
\usepackage{textcomp}
\usepackage{enumitem}
\usepackage[table]{xcolor}
\def\BibTeX{{\rm B\kern-.05em{\sc i\kern-.025em b}\kern-.08em
    T\kern-.1667em\lower.7ex\hbox{E}\kern-.125emX}}

\usepackage{balance}  

\usepackage[]{graphicx}
\graphicspath{{./figures/}}
\DeclareGraphicsExtensions{.pdf,.jpeg,.jpg,.png, .jpg, .svg}

\usepackage{booktabs} 
\newcommand\cparagraph[1]{\vspace{1.01mm}\noindent\textit{#1}}
\newcommand\bparagraph[1]{\vspace{1.01mm}\noindent\textbf{#1}}

\usepackage{caption}
\usepackage{subcaption}

\usepackage{multirow}
\usepackage{tabularx}

\usepackage{titlesec}

\usepackage{array}
\usepackage{caption}
\usepackage{subcaption}
\usepackage{multirow}
\usepackage{tabularx}
\usepackage{soul}

\usepackage[hyphens]{url}
\usepackage[para]{footmisc}

\usepackage[draft]{todonotes}

\begin{document}

\title{Life, the Metaverse and Everything: An Overview of Privacy, Ethics, and Governance in Metaverse}


\author{\IEEEauthorblockN{ Carlos Bermejo Fernandez}
\IEEEauthorblockA{\textit{Hong Kong University of Science and} \\
\textit{Technology}\\
csbermejo@ust.hk}
\and
\IEEEauthorblockN{Pan Hui}
\IEEEauthorblockA{\textit{Hong Kong University of Science and} \\
\textit{Technology} \\
\textit{University of Helsinki}\\
panhui@ust.hk}
}

\maketitle

\begin{abstract}
The metaverse is expected to be the next major evolution phase of the internet. The metaverse will have an impact on human society, production, and life. In this work, we analyze the current trends and challenges that building such a virtual environment will face. We focus on three major pillars to guide the development of the metaverse: privacy, governance, and ethical design, to guide the development of the metaverse. Finally, we propose a preliminary modular-based framework for an ethical design of the metaverse.
\end{abstract}

\begin{IEEEkeywords}
metaverse, virtual worlds, extended reality, ethics, governance, privacy
\end{IEEEkeywords}

\section{Introduction}~\label{sec:introduction}
\begin{figure}[t]
    \centering
    \includegraphics[width=.9\linewidth]{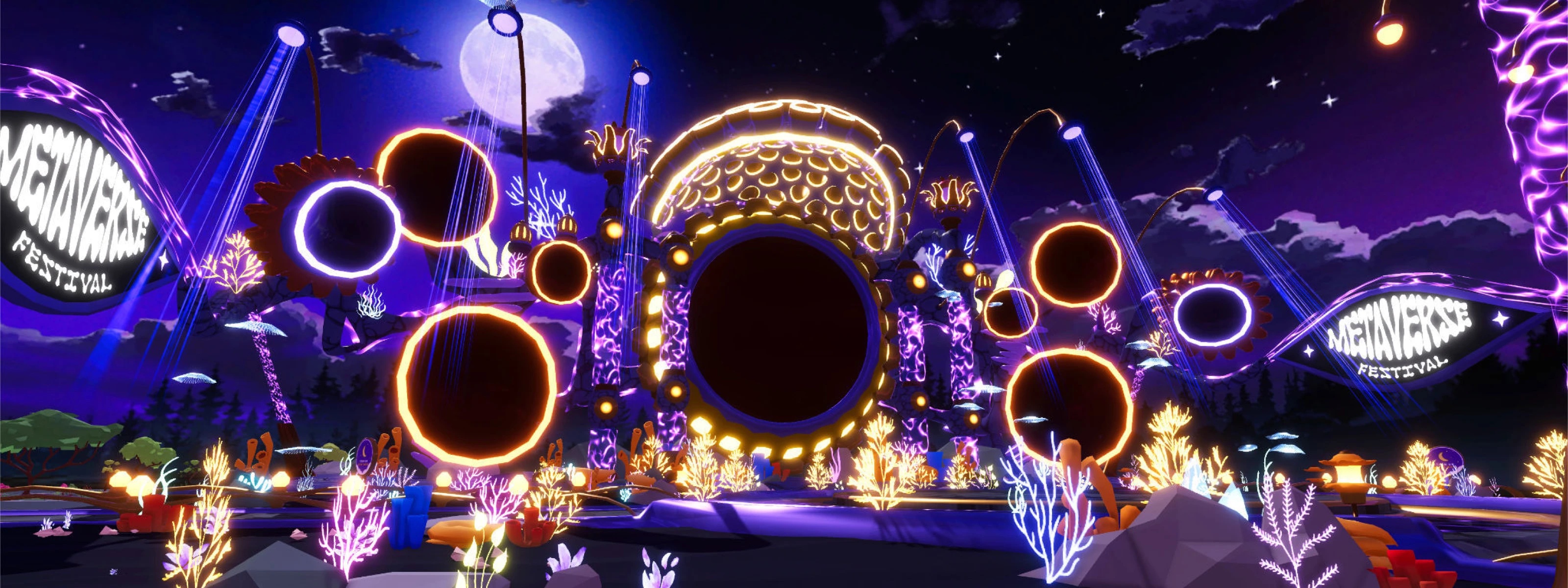}
    \caption{Metaverse festival. Image captured in Decentraland.} 
    \label{fig:intro:metaverse}
\end{figure}

The metaverse has recently increased importance in the web space~\cite{lee2021all}. Online platforms such as 
Decentraland\footnote{\url{https://decentraland.org}} and the Sandbox\footnote{\url{https://www.sandbox.game/en/}} showcase the potential of the metaverse  the first deployed virtual worlds using decentralized tools (e.g., Blockchain), see Figure~\ref{fig:intro:metaverse}. There are also several platforms along this path that contributed to the recent interest in the metaverse, such as Second Life\footnote{\url{https://secondlife.com}}, Minecraft\footnote{\url{https://www.minecraft.net/en-us}}, and Roblox\footnote{\url{https://www.roblox.com}}; and companies, such as Niantic\footnote{\url{https://nianticlabs.com/en/}},  Microsoft (with Mesh\footnote{\url{https://www.microsoft.com/en-us/mesh}}) and recently Meta (before formerly know as Facebook). 

The metaverse will significantly impact our society and our own lives. We are already seeing a change in the trading of virtual assets online and in online games, where users can create and trade digital assets such as accessories for avatars. However, there are still several challenges in creating a metaverse, such as privacy, ethics, and governance. First, the technologies used to create the metaverse introduce new ethical and privacy dilemmas. The continuous sensing of devices to provide more realistic and immersive experiences can be a threat to users' privacy, security and even their safety~\cite{roesner2014security}. The biometrical information such as gaze, gait, heart rate shows important aspects of users' psyche~\cite{renaud2002measuring}. Second, the metaverse can be seen as a microcosmos of our society~\cite{lee2021all}. The governance of such virtual massive worlds presents challenges to regulating the behaviour of users~\cite{olnes2017blockchain}. Therefore, the metaverse requires regulations and policies to manage the platform and its member.

For example, the online virtual worlds, such as Horizons by Meta, face some of the above issues. The misconduct of avatars uses the virtual world of the metaverse as a channel to sexual harass other avatars\footnote{\url{https://www.bbc.com/news/technology-60247542}}. These behaviours raise issues about the punishment of misbehaviour and which regulations and policies the metaverse should have. How will the metaverse regulate the killing of an avatar by another platform member? Will the metaverse follow current policies according to local governments? 


In this work, we discuss the significant challenges that the metaverse will face in terms of security and privacy, ethics, and governance. Moreover, we include trends and approaches that current online worlds are implementing in their solutions and research paths to enable a more sustainable metaverse. We leave open questions to the community regarding critical aspects of the governance and design for an accessible and inclusive metaverse. Finally, we describe a preliminary modular-based approach for an ethical design of the metaverse.

\section{Privacy in the Metaverse}~\label{sec:privacy_security}
Metaverse uses data collected from the real world to provide immersive experiences. Sensors attached to users (e.g., gyroscope to track their head movements) can realistically control their avatar. Moreover, the metaverse also opens new challenges in its massive virtual worlds where users can also be subject to privacy attacks such as eavesdropping by other platform users. In this section, we present the main challenges that developers, designers, practitioners, regulators, and users will face in creating the metaverse.

\subsection{Privacy at sensory level}

Extended Reality (XR) devices enable a more immersive, realistic, and better metaverse experience. XR devices can capture a vast amount of information from biometrical data of users to spatial data, including surroundings such as bystanders' physical space (rooms)~\cite{de2019security,guzman2021unravelling}. As stated by previous works~\cite{guzman2021unravelling,de2019security,roesner2014security}, XR technologies present several privacy and security threats to users and bystanders. These technologies generally use sensors to scan and monitor the users' surroundings~\cite{guzman2021unravelling}. These scans can collect information that might be sensible to users and bystanders that are in the coverage zone of the monitoring~\cite{de2019security,roesner2014security}. 
Head-mounted displays (HMDs) commonly used to display the metaverse can collect some biometric data (head movement, eye tracking) that are non-obvious to users. For example, gaze data can give away users' sexual preferences~\cite{renaud2002measuring}. The collected biometric data puts the most personal aspects of our psyche at risk. Therefore, these devices should treat the data according to some principles that protect the users' privacy.

\begin{figure}
    \centering
    \includegraphics[width=.9\linewidth]{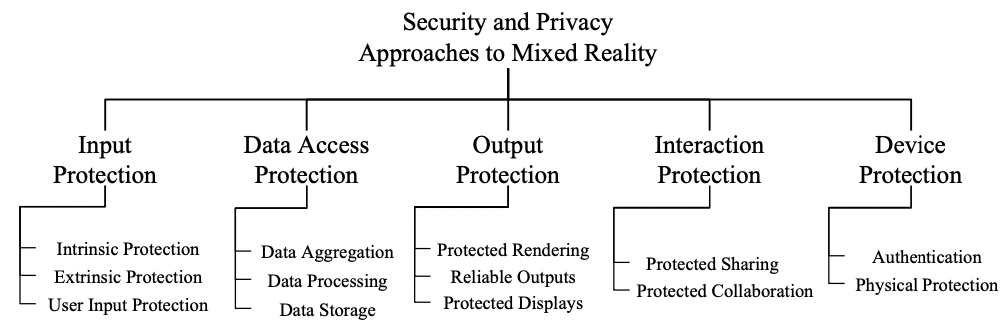}
    \caption{Data-centric perspective to protect the privacy, by De Guzman~\textit{et al.}~\cite{de2019security}.}  
    \label{fig:privacy:data_input}
\end{figure}

\bparagraph{Solutions.} Several works~\cite{de2019security,lebeck2017securing,hu2021lenscap} designed different solutions to protect users privacy in such scenarios. 
These works~\cite{de2019security,lebeck2017securing} propose frameworks to control the privacy of data by securing the input, see Figure~\ref{fig:privacy:data_input}. These frameworks allow users and developers to fine-control the privacy of the input data (from different sensors). This fine-control of collected data can be managed by privacy-enhancing technologies (PETs) that obfuscate any sensible data from the sensors before being shared with cloud services (e.g., online games, metaverse). Although these solutions that control the data collection and sharing with external entities (e.g., cloud servers) still require all parties to create the metaverse (manufacturers, systems, frameworks).

\subsection{Privacy of behaviour and communications}

The metaverse can be seen as a microcosmos of our physical reality. Users can interact with other virtual assets and avatars. The relation and social interactions can be valuable to infer users' habits, activities, and choices in the metaverse. Similarly to the biometric data, this information can tell the users' psyches. Moreover, the metadata inherent in any social interaction with other avatars (e.g., conversations, reactions) presents privacy risks to users. This information could be helpful to track and regulate the behaviour of users. Who is in control of all this information? 

\bparagraph{Solutions.} We can foresee that users can use secondary avatars to obfuscate their real avatar and any possible data that can leak information about users' demographics, cultural, and economic background. These secondary avatars allow users to hide their real behaviour in the metaverse. For example, a user can use a secondary avatar (e.g., clone) when she/he wants to hide their actions in the metaverse~\cite{falchuk2018social}. Other avatars in the metaverse cannot recognise the real owner of this secondary avatar and, therefore, cannot infer any behavioural information about the users.

Users of the metaverse should also have some configurable options to manage their personal space in the virtual world. For example, privacy bubbles restrict visual access with other avatars outside the bubble. Facebook (current `Meta') implemented similar options in their social platform Horizons.

\subsection{Safety of users and bystanders}

Besides the security challenges involved in the protection of information collected while being shared (e.g., sensors, users' action in the metaverse) and against tampering, other security factors might affect users, such as safety~\cite{dao2021bad,tseng2022dark}. The current HMDs that are used to display the metaverse can occlude the physical world and the ability of users to detect nearby objects, increasing the risk of falling~\cite{dao2021bad,tseng2022dark}. As described in previous sections, the metaverse can also be visualized as an augmentation of the physical world using MR. Therefore, the risks of occluding dangers such as cars while the user is walking on the streets.

\bparagraph{Solutions.} In~\cite{langbehn2018shadow}, the authors propose the visualization of real users (individuals located in the same room as the user) as virtual (`shadow') avatars to avoid collisions in multi-user VR experiences. Redirecting users' walking while disrupting their immersion in the virtual world reduces the collision with physical objects in their surroundings~\cite{bachmann2019multi,sun2018towards}. 
Other solutions could use sensors implemented in the headsets to spatially scan the room and display the physical objects in the virtual world in case of possible collisions. 
These solutions should still require the same privacy protections as other sensors and not incur the collection of more sensible data.

\subsection{Open challenges}

Despite all the aforementioned proposed solutions, there is still the need for XR platforms (device manufacturers, systems, frameworks) to develop a consistent privacy protection solution across all entities that conform to the metaverse. 
Future of Privacy Forum\footnote {\url{https://fpf.org}} advocates for the processing of private data on the users' side (e.g., XR device). XR devices that collect sensible data should provide granular control (switches) to manage the input data flows from sensors and provide visual cues (e.g., LED in the device) when personal data is collected or transmitted. As we have observed, despite the inclusion of fundamental privacy protection tools in virtual worlds (privacy bubble in Horizon Worlds), users are either not fully aware of them or do not know how to use them. The privacy regulations and practices should be transparent and clear to all members of the metaverse. Following the guidelines of the XR Privacy Summit, all the players involved in creating and managing the metaverse should adopt some form of institutional review board (IRB) model in their organisms. Companies must deploy a mix of technical solutions and policies while complying with the law to earn widespread consumer trust and adoption.

We have an example with local regulators such as the California Consumer Privacy Act (CCPA) and the General Data Regulation Protection (GDPR) that target the protection of individuals in monitoring environments. The purpose of these regulations is analogous in protecting users' data, despite coming from different local laws. Using a modular-based framework to construct the privacy regulation protections will allow the metaverse to adapt to local authorities' specifications and provide a homogeneous policy to protect users' privacy. The metaverse should advocate for PETs and in-sensor data processing practices. A distributed ledger (Blockchain) can register any party's data collection and processing activities in the metaverse. Finally, the metaverse should guarantee no data monopoly from any parties in the data collection practices.


\section{Governance in the Metaverse}~\label{sec:governance}
Before we analyze the possibilities of governance in the metaverse, we can study the current responses that online social media and gaming platforms are following to govern them and their societal impacts on users~\cite{tarleton_custodians_2018,haimson2016constructing}. Users of these platforms face issues of misbehaviour, spam, harassment, and conflicts with other users of the platform~\cite{schneider2021modular}. Online communities present several challenges when these grow in size and moderators (initially other members of the community) cannot keep up with the demand of comments and misbehaviour of the community members. In the case of social networks such as Facebook and Twitter, automation tools have been included to control misbehaviour (e.g., banning inappropriate posts). These platforms also rely on the report of other members to manage the spam and wrong posting of other members. Similarly, players of massively multiplayer online games forms communities to self-govern the misbehaviour of other players (e.g., stealing digital assets of other players, spamming)~\cite{humphreys2008ruling}. This section illustrates how current approaches from online social and multi-user gaming platforms could be used to regulate the (`limitless') metaverse.



\subsection{Code and rules}

We can see the software code of the metaverse as an analogy to our physical laws of nature, where code can constrain the shape of the metaverse. Code shapes online environments and the behaviour of users~\cite{lessig_code1999}. Developers and companies can decide what features will be included in the online platform. In the case of open-source approaches, developers can take social decisions to achieve the respective goals of their idea of metaverse~\cite{humphreys2008ruling}.  

Code rules can also influence the social behaviour of users in the metaverse. The choices that developers make during the metaverse development can impact how users interact with the platform~\cite{lessig_code1999}. For example, developers configure a privacy-bubble mode where users can set their private space (bubble) and restrict access (e.g., interactions such as chat). This privacy mode can change how people behave in the metaverse as they always have an available tool to restrict access to other avatars.

Despite the `rules of the code', users usually find new ways to interact with online platforms outside the rules defined by the code~\cite{humphreys2008ruling}. Hacking is a good example where modifications in the original code enable new virtual worlds, including new creation tools, interactions, and overall experience. For example, online games such as Grand Theft Auto and Skyrim have been the target of mods in their online version, which completely changes the gaming experience~\cite{humphreys2008ruling}. However, we cannot govern all possible social behaviour on online platforms by using code.

\subsection{Blockchain and decentralized autonomous organizations}

Blockchain is a transformation driver for online platforms as it enables the creation of information integrity and smart contracts~\cite{olnes2017blockchain}. Blockchain has been seen as a disruptive technology that supports information exchange and transactions that require authentication and trust~\cite{olnes2017blockchain}. This technology shows the potential to provide benefits for governance in the metaverse.

Decentralized autonomous organizations (DAOs) are based on Blockchain and smart contract technologies. These decentralized organizations allow online platforms, such as the metaverse for global collaboration and coordination. Generally, DAOs are usually flat and fully democratized, where each member can participate in the voting system to implement any changes in the platform. The system can also automatically handle services, such as selling a property asset in the metaverse, while being transparent and fully accessible to any metaverse user. We can see a real-world example of DAOs in online platforms such as Decentraland and the Sandbox, aiming to provide metaverse virtual worlds. In these platforms, users can contribute to the decision-making. However, DAOs can face several scalability issues and integration with the metaverse community. The flat-based design of several DAOs can hinder the members' involvement in the decision-making process as the number of voting sessions can become cumbersome. The algorithms embedded in autonomous decisions can strongly impact the overall metaverse. Open source approaches to building the metaverse are encouraged. However, understanding the algorithms can be more challenging for the public, and therefore auditing systems within the platform should be available.

Finally, we should highlight that these computation approaches cannot capture the fullness of human governance practices but can facilitate the non-computational process such as voting or content moderation in the metaverse.

\subsection{Modular governance}

Governance in online platforms is a high-stakes challenge that yet has few basic features of offline (physical) governance legacies~\cite{schneider2021modular}. The authors~\cite{schneider2021modular} propose a modular bottom-up governance approach for online platforms in this work. This modularity can enable the development of portable tools that can be adapted to different platforms and use cases. The governance layer should include a broad spectrum of processes (juries, formal debates) and interact with other governance systems. An example of governance systems that follow the above features is decentralized autonomous organizations (DAOs). These governance systems allow users to be actively involved in decision-making processes (e.g., Decentraland, Sandbox).

\subsection{Online platforms and the social good}

Online environments can also be seen as viable sites to shape the behaviour of youth positively~\cite{tekinbacs2021designing}. The findings from a 2-year (in-the-wild) study on (custom) Minecraft server suggest the shaping the visions of youth's (8-13 years old) capacity for ownership and control mechanisms to manage the culture and climate of the online community. The study also highlights the importance of moderators (adult intervention in this case) to solve some problems in the custom community. The authors also suggest that online platforms should consider tools to deal with players' misbehaviour (i.e., punitive approaches) and tools for encouraging positive behaviours (i.e., preventive approaches). They also propose incentive mechanisms to promote positive behaviour and restrain negative players. These incentive systems can also encourage collaboration, shared planning, and teamwork. Finally, the metaverse should consider implementing tools (e.g., DAOs) to allow users to engage in discussion, reflection, and decision-making processes.

We can see the importance of community governance in influencing and managing online behaviours (including moderation and conflict resolution).

\subsection{Open Challenges}

We should ask ourselves who will govern the metaverse? Will these governments be based on current local laws? The modular approaches offer a solution to adapt the regulations and control of the metaverse according to the scenario. For example, if the metaverse is required to follow the local rules, the modules will swap accordingly. Then, the question is how the users from other geographical locations will be treated and how these local rules can be applied to the metaverse and aim for a global virtual world. We could end up with a version of the metaverse with frontiers, in which the regulations are applied differently. Ideally, as we will see in the last section, the metaverse should be inclusive and outside of local regulations that diminish users' freedom.

We can foresee that the described tools and approaches could shape the metaverse and accommodate the necessary regulations accordingly.

\section{Ethical design for the Metaverse}~\label{sec:ethical_design}
The metaverse has the potential to open revolutionize our current society, where new channels to express ourselves and interact with others without any limits (location, time, race, gender). This section illustrates the positive impact of the metaverse on our society and our proposed ethical design.

\subsection{Creation in the metaverse}

The creation process will be a fundamental asset in the metaverse. As we have already seen with current platforms Decentraland and the Sandbox, the creation of digital assets has opened a new market for innovation, monetary income, and jobs. Non-fungible tokens are unique digital tokens (using Blockchain technologies) that represent the ownership of a particular asset, such as digital artwork and collectables. NFTs are a one-to-one mapping between an owner (represented by a crypto wallet address) and the asset referencing the NFT (usually by a uniform resource identifier, URI). NFTs replicate the properties of physical objects such as scarcity and uniqueness~\cite{sharma2022s}. For example, Decentraland uses NFTs to manage the game's virtual lands and other digital assets, such as clothing accessories for the avatars. 


Besides the markets that trade NFTs, we can see a new revolution in online platforms such as play-to-earn games, where NFTs-powered games such as Axie Infinite\footnote{\url{https://axieinfinity.com/}} (a Pokemon-style monster-battling) allow players to earn money while they play, they can sell their improved monster. Other models that virtual worlds are including in their platforms (e.g., Sandbox) is the create-to-earn where users of the platform can contribute to its construction while selling their created digital assets.

The lack of centralized authorities empowers the NFTs as a democratization content creation and commercialization tool. 
However, these democratization tools to lower the barriers to creating and sharing content also allow scammers and malicious content creators to take advantage of the system to sell copies or low-quality NFTs. 
Several trading platforms of NFT are using `invite-only' policies to allow only a specific group of creators in their platforms. This kind of policy diminishes the advantages of NFTs as an open-access content creation tool. 
A possible solution can be seen in using DAOs and users of the platform to implement a reputation-based system where everyone can vote and enforce norms to keep the quality of NFTs and reduce scams. Also, other works~\cite{kou2020mediating,chandrasekharan2019crossmod,sharma2022s} highlight the use of moderation practices using AI technologies to reduce toxic behaviours in games, online forums, and content creation. These AI-based and cross-modality solutions include users of the platform and AI tools to provide a scalable moderation framework applied in the metaverse. Further research is needed to find a balance between NFTs and quality control.

\cparagraph{Digital twins.} We can define digital twins as virtual objects that are created to reflect physical objects, including the appearance and physical behaviour (of the real world). Digital twins, due to their physical - virtual synchronization, will allow users of the metaverse to view themselves not only in the realms of virtual worlds in VR but in other paradigms such as mixed reality (MR), where virtual and physical are merged. We can also foresee that actions of users in the physical world, such as travelling taking a photo, will feed the metaverse and the avatars in it (e.g., displaying the photo in the virtual world). The metaverse will be then an evolving world that is synchronized with the physical one. There are still some challenging regarding ownership of digital twins. The most straightforward approach to protecting digital twins' authenticity and origin is using a digital ledger such as Blockchain.

\subsection{The social metaverse}

The metaverse can show a positive impact on social good in terms of accessibility, diversity, equality, and humanity~\cite{duan2021metaverse}. 

\bparagraph{Accessibility.} The metaverse can enable global collaboration despite the geographical distances. Moreover, this digital world can also provide accessibility for social events such as concerts. The metaverse can enable many social events that are not possible physically - for example, concerts with millions of people worldwide. For example, in 2020, UC Berkeley held its graduation ceremony in Minecraft. 

\bparagraph{Diversity.} The limitations of the physical world can be lifted when using the metaverse. The metaverse can have unlimited spaces and virtual worlds. Moreover, there are no limits to our activities in the metaverse. We can imagine the metaverse as a place where users can display their artwork, socialise, play, learn, and more. 

\bparagraph{Equality.} The metaverse can be seen as an equaliser where gender, race, disability, and social status are eliminated. Users can customise their avatars, where their imagination is the limit (e.g., they can be a cat). This feature will allow the metaverse to build a fair and more sustainable society in the virtual world.

\bparagraph{Humanity.} The metaverse can be a door to cultural communications and protections. For example, the metaverse can be the platform to preserve and restore art pieces. However, this can open challenges on how the platform will preserve different cultural and artistic forms. The metaverse can create unlimited communities with different beliefs and thoughts, allowing everyone to be part of the metaverse.

\bparagraph{Trust.} However, as we have seen in the introduction, the anonymity of online communities can raise concerns about how we transmit information in the metaverse. We can see raising concerns with fake-news reports and the reflections that the `bad' internet can have in transmitting information and, therefore, knowledge. In the physical world, humans traditionally gain knowledge through rule-based analyses. Misinformation (e.g., fake news) affects the culpable and the gullible that share misinformation despite good intentions (e.g., inform, warn). In the metaverse, testimonies and trust will play an even more critical role, as in many cases, we will not have a real person telling the testimony but her/his avatar. Several experts explore additional reaction principles (e.g., punishment) to show why truth denial is worse than false belief. Incentive systems to share trust among avatars will be key functionality to reduce the sharing of misinformation.

Organizations such as XR Safety Initiative (XRSI)\footnote{\url{https://xrsi.org}} and XR Association\footnote{\url{https://xra.org}} promote the responsible design and adoption of XR solutions (including the metaverse). These organisations can encourage companies and institutions to follow ethical designs in their metaverse implementations (e.g., XR devices).

\subsection{Modular ethical design}

Human-Centered Design (HCD) is a philosophy developed by Don Norman (among others)~\cite{norman2005human}. We consider this design approach for the metaverse as it requires the active involvement of users in the design process and decision-making. As we have seen, the metaverse will significantly impact human society, production, and life. A modular-based metaverse architecture will allow adapting to the specifications and requirements of such a worldwide platform. Therefore, our preliminary approach aims to involve every necessary member (developers, regulators, users, content creators) in the design and implementation of the metaverse.

\begin{figure}
    \centering
    \includegraphics[width=.9\linewidth]{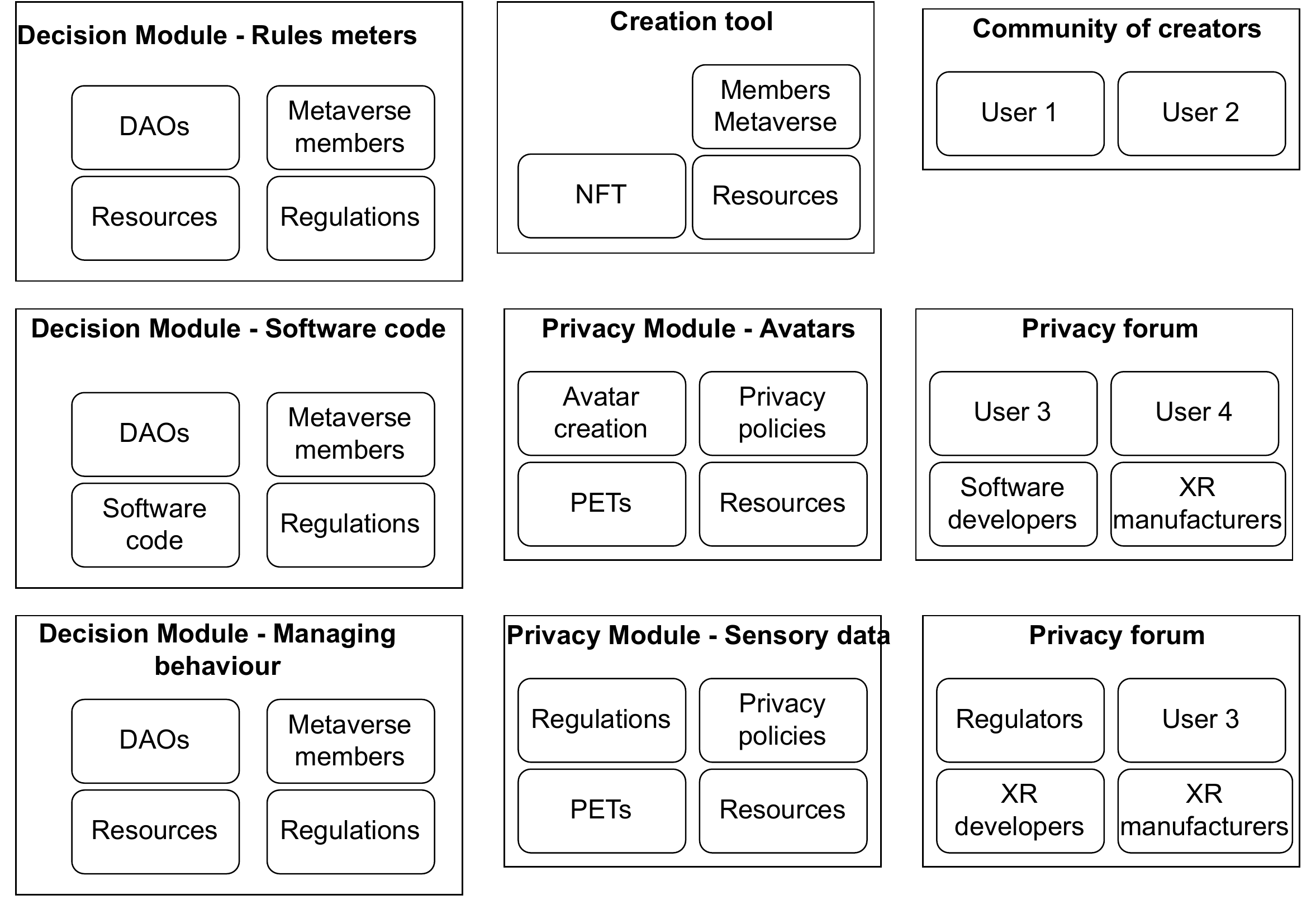}
    \caption{An example of a modular-based metaverse architecture motivated the 'Ethical Hierarchy of Needs', where each module is interchangeable.} 
    \label{fig:modular:metaverse}
\end{figure}

Figure~\ref{fig:modular:metaverse} illustrates several examples of modules that will realize a specific task. For example, the decision-making module will involve members, regulators, and software developers. We believe that DAOs can solve the scalability problems when those are spread across (modular approach) different features of the metaverse. We can see these modules as a federated approach. These modules can take independent decisions such as the reaction to misbehaviour, but are still connected to other decision modules, resources, and policies. As we have discussed in previous sections, these decision algorithms should be transparent to every member of the metaverse. The changes in the metaverse will also involve code and hardware implementations as they can affect the privacy and security of users. Moreover, some default privacy protection rules should be implemented to protect the users' information and reduce the chances of data monopoly, as is happening with our current internet. These decisions will follow (if necessary) local government regulations and adapt to any future change.

The metaverse will require an efficient system to share trust across all layers. A reputation-based system under the Blockchain will enable the metaverse with a tool to counterbalanced attacks during decision-making processes and limit the spread of misinformation (e.g., fake news). Machine learning or artificial intelligence tools to automate processes in the metaverse should be a subject of debate and follow current GDPR policies to explainability and transparency.

We illustrate the alignments of the metaverse with the `Ethical Hierarchy of Needs' (licensed under CC BY 4.0) for an ethical design:

\bparagraph{Human rights.} The metaverse should be accessible, diverse, and inclusive where every individual is welcome. Privacy is an intrinsic right of users, which they can protect using the available tools (PETs, avatar creation). All the modules are interchangeable and will use the current technologies such as Blockchain and DAOs to manage any decision taken in the metaverse (e.g., change in current policies about users' misbehaviour). All the active parts of the metaverse (including code) should be transparent and understandable to any platform member. This can reduce the complexity while voting (using DAOs) in the decision-making processes.

\bparagraph{Human effort.} The metaverse will include a reputation-based system that will be inherently attached to users and will be managed by Blockchain and DAOs. This reputation system will allow users to report malicious users' misbehaviour and malpractice while voting using DAOs. Users of the metaverse, developers, content creators, and regulators will be involved in the decision-making process of the metaverse.

\bparagraph{Human Experience.} The metaverse will be an immersive experience that will enable new ways of social interaction content creation beyond our current real-world boundaries (e.g., regulations, geographical location). The metaverse will be an accessible and inclusive virtual world due to the possibility of creating a myriad of avatars and communities.

\bparagraph{Limitations.} We propose a modular approach for the metaverse and an approach to enable an ethical design. However, we have not proven that such tools will lead to the involvement of all parties (e.g., users of the metaverse) in fair decisions towards an ethical goal. Still, examples in current platforms such as Decentraland, the Sandbox, and MMOGs have shown the metaverse's potential as a social good and the participation and collaboration of users in the decision-making (using DAOs in case of the first two).


\section{Conclusion}~\label{sec:conclusion}
This paper presents an overview of the main challenges that the metaverse will face regarding privacy, governance, and ethics. We also show a preliminary step towards an ethical design for the metaverse. Still, much research and experimentation remain in order before the metaverse becomes the utopia that many foresee.


\bibliographystyle{IEEEtran}
\bibliography{socialmeta_ethics_metaverse} 



\end{document}